\documentclass[showpacs,10pt,twocolumn,prb]{revtex4-1}

\usepackage{amsmath}
\usepackage{amssymb}
\usepackage{graphicx}
\usepackage{amssymb}
\usepackage{graphics}
\usepackage{epsfig}
\usepackage{CJK}
\usepackage{color}

\begin{document}

\begin{CJK*}{GBK}{Song}
\title{Anisotropic magnetocaloric effect in single crystals of CrI$_3$}
\author{Yu Liu and C. Petrovic}
\affiliation{Condensed Matter Physics and Materials Science Department, Brookhaven National Laboratory, Upton, New York 11973, USA}
\date{\today}

\begin{abstract}
We report a systematic investigation of dc magnetization and ac susceptibility, as well as anisotropic magnetocaloric effect in bulk CrI$_3$ single crystals. A second-stage magnetic transition was observed just below the Curie temperature $T_c$, indicating a two-step magnetic ordering. The low temperature thermal demagnetization could be well fitted by the spin-wave model rather than the single-particle model, confirming its localized magnetism. The maximum magnetic entropy change $-\Delta S_M^{max} \sim 5.65$ J kg$^{-1}$ K$^{-1}$ and the corresponding adiabatic temperature change $\Delta T_{ad} \sim 2.34$ K are achieved from heat capacity analysis with the magnetic field up to 9 T. Anisotropy of $\Delta S_M(T,H)$ was further investigated by isothermal magnetization, showing that the difference of $-\Delta S_M^{max}$ between the $ab$ plane and the $c$ axis reaches a maximum value $\sim$ 1.56 J kg$^{-1}$ K$^{-1}$ with the field change of 5 T. With the scaling analysis of $\Delta S_M$, the rescaled $\Delta S_M(T,H)$ curves collapse onto a universal curve, indicating a second-order type of the magnetic transition. Furthermore, the $-\Delta S_M^{max}$ follows the power law of $H^n$ with $n = 0.64(1)$, and the relative cooling power RCP depends on $H^m$ with $m = 1.12(1)$.

\end{abstract}
\maketitle
\end{CJK*}

\section{INTRODUCTION}

Layered intrinsically ferromagnetic (FM) semiconductors hold great promise for both fundamental physics and future applications in nano-spintronics.\cite{McGuire0, McGuire, Huang, Gong, Seyler} For instance, Cr$_2$X$_2$Te$_6$ (X = Si, Ge) and CrX$_3$ (X = Cl, Br, I) have recently attracted wide attention as promising candidates for long-range magnetism in monolayer.\cite{Gong, Lin, Zhuang}

Bulk Cr$_2$X$_2$Te$_6$ (X = Si, Ge) shows FM order with the Curie temperature ($T_c$) of 32 K for Cr$_2$Si$_2$Te$_6$ and 61 K for Cr$_2$Ge$_2$Te$_6$, respectively.\cite{Ouvrard, Carteaux1, Carteaux2, Casto, Zhang} Scanning magneto-optic Kerr microscopy experiment shows that the $T_c$ monotonically decreases with decreasing thickness of Cr$_2$Ge$_2$Te$_6$, from bulk of 68 K to bilayer of 30 K.\cite{Gong} Similarly, bulk CrI$_3$ shows FM with $T_c$ of 61 K, and the magnetism can persist in mechanically exfoliated monolayer with $T_c$ of 45 K.\cite{Huang} The magnetism in CrI$_3$ is intriguingly layer-dependent, from FM in the monolayer, to antiferromagnetic (AFM) in the bilayer, and back to FM in the trilayer,\cite{Huang} providing great opportunities for designing magneto-optoelectronic devices. A rich phase diagram, including in-plane AFM, off-plane FM, and in-plane FM, is further predicted by applying lateral strain and/or charge doping.\cite{Zheng} Besides, the magneto-transport measurement on the thin exfoliated CrI$_3$ reveals a tunneling magnetoresistance as large as 10,000$\%$, exhibiting multiple transitions to the different magnetic states.\cite{Wang, SongTC} The magnetism in monolayer and/or bilayer CrI$_3$ can also be controlled by electrostatic doping using a dual-gate field-effect device.\cite{Jiang, Huang1}

Since the layered van der Waals magnetic materials may exibit magnetocrystalline anisotropy, in this paper, we study magnetocaloric effect by heat capacity and magnetization measurements around $T_c$.\cite{ZhangWB,VerchenkoVY} Isothermal magnetic entropy change $\Delta S_M(T,H)$ can be well scaled into a universal curve independent on temperature and field, indicating the magnetic transition is of a second-order type. Moreover, the $\Delta S_M(T,H)$ follows the power law of $H^n$ with $n = dln\mid \Delta S_M \mid /dln(H)$. The temperature dependence of $n$ values reaches minimum at $T \sim 60$ K, at $T_c$ of bulk CrI$_3$.

\section{EXPERIMENTAL DETAILS}

Single crystals of CrI$_3$ were grown by chemical vapor transport (CVT) method and characterized as described previously.\cite{LIUYU} The heat capacity was measured in Quantum Design PPMS-9 system with field up to 9 T. Several crystals with mass of 5.6 mg were used in heat capacity measurement. The magnetization data as a function of temperature and field were collected using Quantum Design MPMS-XL5 system in temperature range from 10 to 100 K with a temperature step of 2 K around $T_c$ and field up to 5 T. One crystal with mass of 1.67 mg was covered with scotch tape on both sides and used in magnetization measurement. The magnetic entropy change $-\Delta S_M$ from the magnetization data was estimated using a Maxwell relation.

\section{RESULTS AND DISCUSSIONS}

\begin{figure}
\centerline{\includegraphics[scale=1]{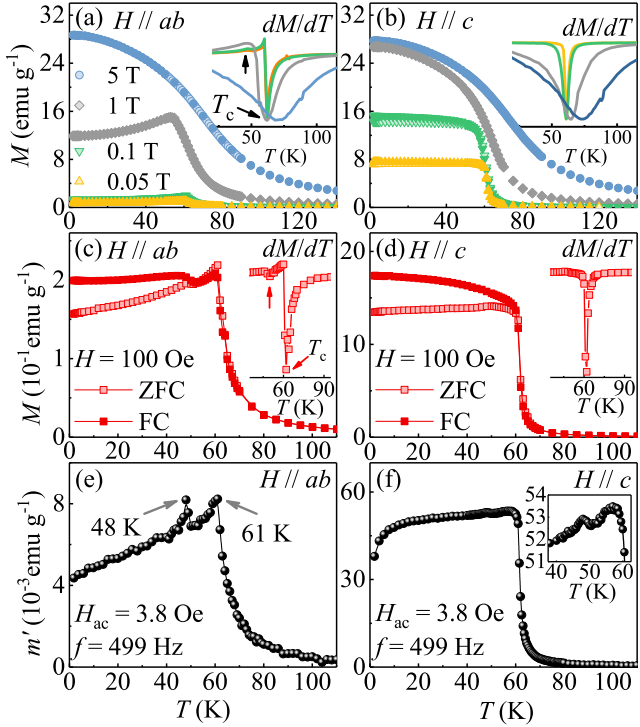}}
\caption{(Color online). Typical temperature dependences of zero-field cooling (ZFC) and field-cooling (FC) magnetizations, $M(T,H)$, of CrI$_3$ measured at the indicated magnetic fields applied (a,c) in the $ab$ plane and (b,d) along the $c$ axis. Insets are the $dM/dT$ curves. Ac susceptibility real part $\chi^\prime(T)$ as a function of temperature measured with oscillated ac field of 3.8 Oe and frequency of 499 Hz applied (e) in the $ab$ plane and (f) the $c$ axis. Inset shows the enlargement region around $T_c$.}
\label{1}
\end{figure}

Figures 1(a) and 1(b) exhibit the temperature dependences of zero-field cooling (ZFC) and field-cooling (FC) magnetizations $M(T,H)$ for bulk CrI$_3$ single crystal measured in the fields ranging from 0.05 to 5 T applied in the $ab$ plane and along the $c$ axis, respectively. An apparent increase in $M(T,H)$ around $T_c$ is observed, which corresponds well to the reported paramagnetic (PM) to FM transition.\cite{McGuire} The $T_c$ is roughly defined by the minima of the $dM/dT$ curves [insets in Figs. 1(a) and 1(b)], which is about 61 K in low fields and increases to 74 K in 5 T. The magnetization is nearly isotropic in 5 T, while in low fields, significant magnetic anisotropy is observed at low temperatures. An additional weak kink was observed below $T_c$ with field in the $ab$ plane. Below $T_c$, the magnetization measured along the $c$ axis saturates at a relatively
low field, indicating that the moments are aligned in this direction. As shown in Figs. 1(c) and 1(d), the splitting of ZFC and FC curves is observed at low field of $H$ = 100 Oe, which originates from the anisotropic FM domain effect. The anomaly below the PM-FM transition is clearly observed. In order to determine the accurate transition temperatures, ac susceptibility was measured at oscillated ac field of 3.8 Oe and frequency of 499 Hz. Two distinct peaks in the real part $\chi^\prime(T)$ in the $ab$ plane [Fig. 1(e)], the PM-FM transition at 61 K and an additional peak at 48 K, as well as the weak anomalies at corresponding temperatures along the $c$ axis [inset in Fig. 1(f)], confirm it is a two-step magnetic ordering. This is in agreement with $d\chi(T)/dT$ data in Ref. 2 where additional anomaly was observed below bulk FM transition at 61 K.

\begin{figure}
\centerline{\includegraphics[scale=0.9]{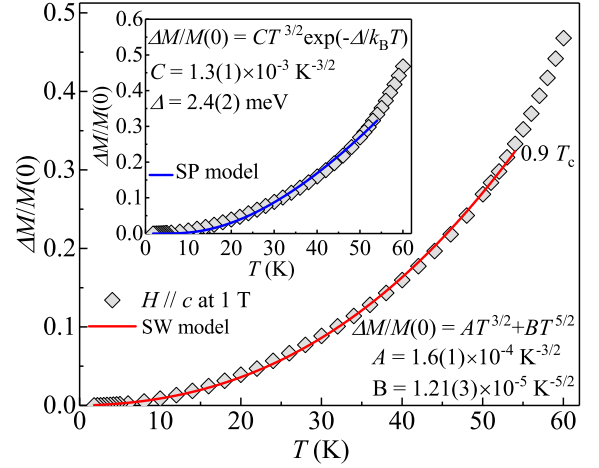}}
\caption{(Color online). Temperature-dependent reduced magnetization of CrI$_3$ fitted using spin-wave (SW) model and (inset) single-particle (SP) model.}
\label{2}
\end{figure}

Figure 2 displays the low temperature thermal demagnetization analysis for CrI$_3$ measured at $H$ = 1 T along the $c$ axis using both spin-wave (SW) and single-particle (SP, inset) models. For a localized moment, the thermal demagnetization at low temperature is generally explained with the spin-wave excitations which follows Bloch equation:\cite{Das, Kaul1, Kaul2}
\begin{equation}
\frac{\Delta M}{M(0)} = \frac{M(0)-M(T)}{M(0)} = AT^{3/2}+BT^{5/2}+...,
\end{equation}
where $A$ and $B$ are the coefficients, $M(0)$ is the magnetization at 0 K. The $T^{3/2}$ term arises due to harmonic contribution and the $T^{5/2}$ term originates from higher order term in spin-wave dispersion relation. While in itinerant or band magnetism where the net moment of system is directly proportional to the displacement energy between spin-up and spin-down subbands, the thermal demagnetization is realized as a result of excitation of electrons from one subband to the other. The single-particle excitation is generally expressed as,\cite{Das}
\begin{equation}
\frac{\Delta M}{M(0)} = \frac{M(0)-M(T)}{M(0)} = CT^{3/2}exp{\frac{-\Delta}{k_BT}},
\end{equation}
where $C$ is the coefficient, $\Delta$ is the energy gap between the top of full subband and the Fermi level and $k_B$ is the Boltzmann constant. Usually the $M(0)$ can be estimated from extrapolation of $M(T)$ data. As shown in Fig. 2, the SW model gives a better fitting result than the SP model up to 0.9 $T_c$, suggesting localized magnetism of insulating CrI$_3$. The fitting yields $A = 1.6(1)\times10^{-4}$ K$^{-3/2}$, $B = 1.21(3)\times10^{-5}$ K$^{-5/2}$, $C = 1.3(1)\times10^{-3}$ K$^{-3/2}$ and $\Delta = 2.4(2)$ meV. It is not surprising that the SP model fails due to intermediate correlation in CrI$_3$. Therefore, a more accurate treatment would be Moriya's self-consistent renormalization theory.\cite{Moriya} It further gives that the magnetic anisotropy in CrI$_3$ comes predominantly from the anisotropic symmetric superexchange via Cr-I-Cr with large spin-orbital coupling rather than the single ion anisotropy.\cite{Lado}

\begin{figure}
\centerline{\includegraphics[scale=0.86]{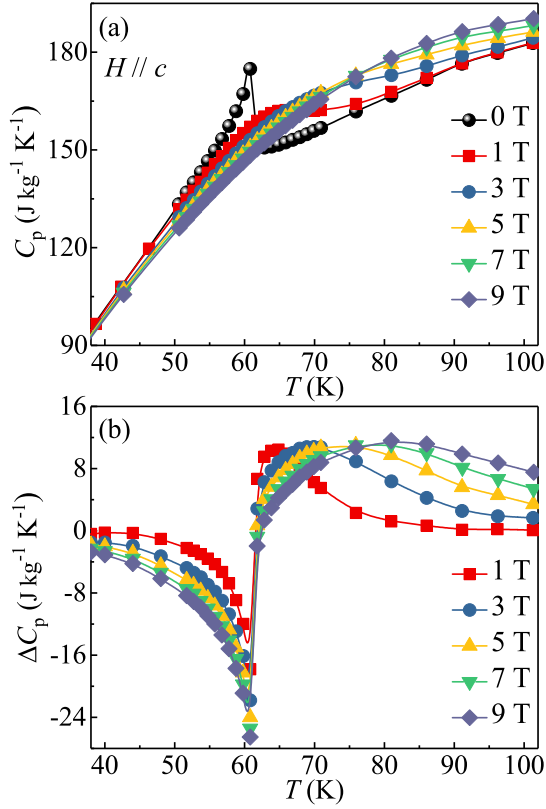}}
\caption{(Color online). Temperature dependence of (a) specific heat $C_p$ and (b) specific heat change $\Delta C_p = C_p(T,H) - C_p(T,0)$ at the indicated magnetic fields.}
\label{3}
\end{figure}

\begin{figure}
\centerline{\includegraphics[scale=0.85]{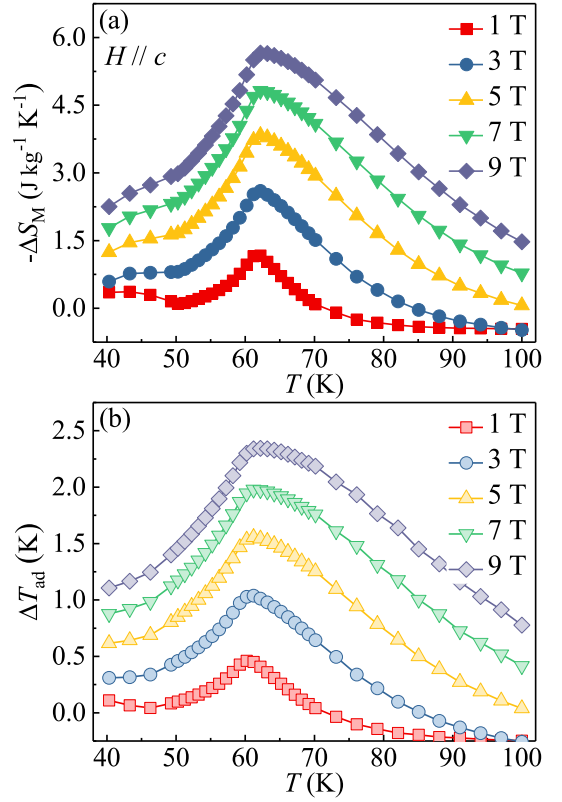}}
\caption{(Color online). Temperature dependences of (a) the magnetic entropy change $-\Delta S_M$ and (b) the adiabatic temperature change $\Delta T_{ad}$ for CrI$_3$ estimated from heat capacity data in different magnetic fields.}
\label{4}
\end{figure}

Temperature dependence of heat capacity $C_p$ for CrI$_3$ measured in various fields along the $c$ axis is presented in Fig. 3(a). A clear $\lambda$-type peak that corresponds to the PM-FM transition is determined to be at $T_c$ = 61 K in zero field, in line with the magnetization data [Figs. 1(e) and 1(f)]. Its height is lowered, broadened and shifts to higher temperatures with increase in magnetic field. The estimated heat capacity change $\Delta C_p = C_p(T,H) - C_p(T,0)$ as a function of temperature in various fields is plotted in Fig. 3(b). Obviously, $\Delta C_p < 0$ for $T < T_c$ and $\Delta C_p > 0$ for $T > T_c$, whilst, it changes sharply from negative to positive at $T_c$, corresponding to the change from FM to PM region. The entropy $S(T,H)$ can be calculated as:
\begin{equation}
S(T,H) = \int_0^T \frac{C_p(T,H)}{T}dT.
\end{equation}
Assuming the electronic and lattice contributions are not field dependent and in an adiabatic process of changing the field, the magnetic entropy change $\Delta S_M$ should be $\Delta S_M(T,H) = S_M(T,H)-S_M(T,0)$. The adiabatic temperature change $\Delta T_{ad}$ caused by the field change can be indirectly determined, $\Delta T_{ad}(T,H) = T(S,H)-T(S,0)$, where $T(S,H)$ and $T(S,0)$ are the temperatures in the field $H \neq 0$ and $H = 0$, respectively, at constant total entropy $S$. Figures 4(a) and 4(b) exhibit the $-\Delta S_M$ and $\Delta T_{ad}$ estimated from heat capacity data as a function of temperature in various fields, both of which exhibit its maximum near the Curie temperature and increase with increasing fields. The maxima of $-\Delta S_M$ and $\Delta T_{ad}$ reach the values of 5.65 J kg$^{-1}$ K$^{-1}$ and 2.34 K, respectively, with the magnetic field of 9 T. Since there is a large magnetic anisotropy in CrI$_3$, it is of interest to calculate the individual magnetic entropy change for the two directions respectively.

\begin{figure}
\centerline{\includegraphics[scale=0.85]{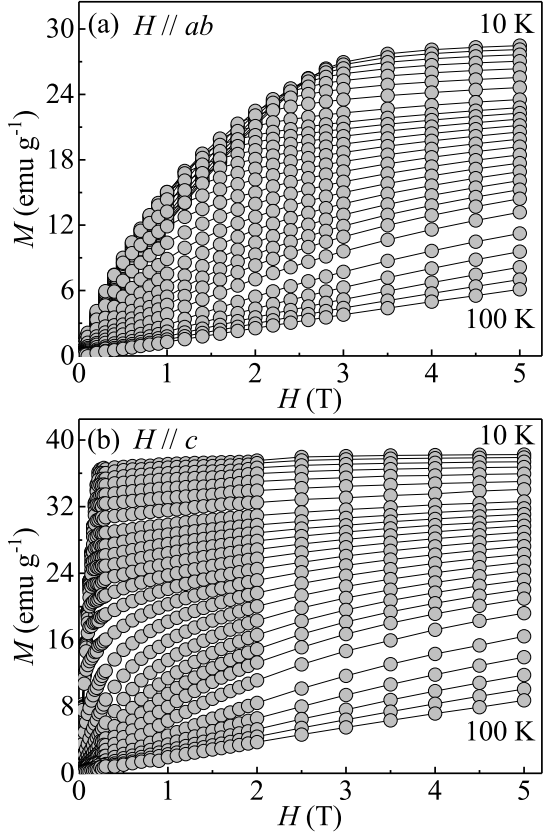}}
\caption{(Color online). Representative magnetization isothermals at various temperatures for (a) the $ab$ plane and (b) the $c$ axis.}
\label{5}
\end{figure}

\begin{figure}
\centerline{\includegraphics[scale=0.868]{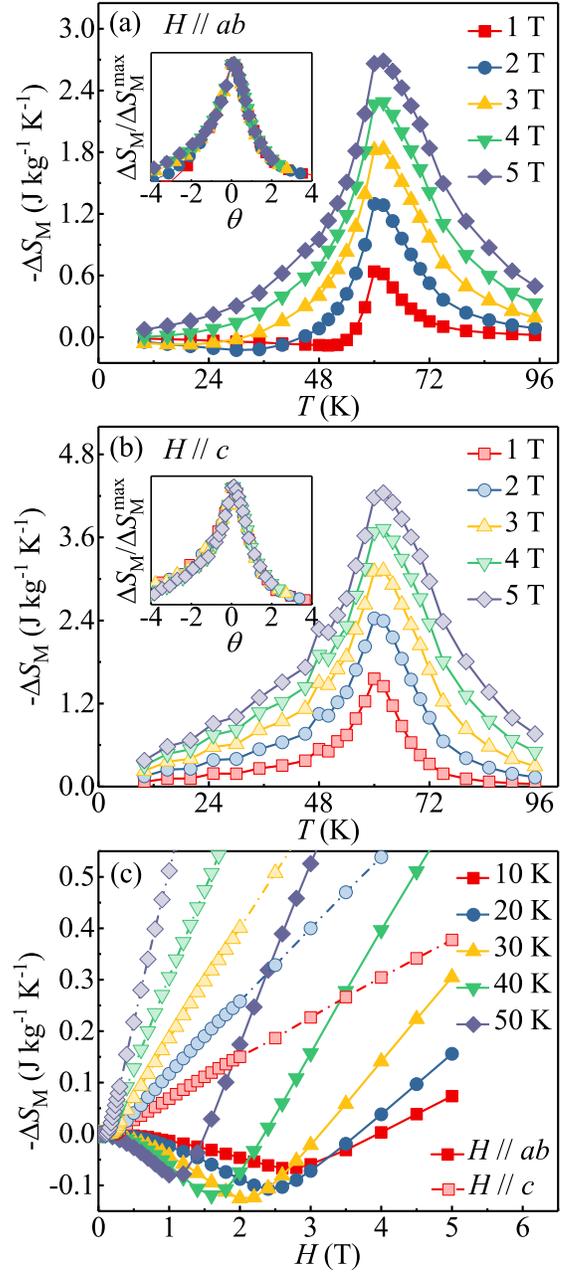}}
\caption{(Color online). Temperature dependence of isothermal magnetic entropy change $-\Delta S_M$ obtained from magnetization measurements at various magnetic fields change (a) in the $ab$ plane and (b) along the $c$ axis. Insets: the normalized $\Delta S_M$ as a function of the rescaled temperature $\theta$. (c) Field dependence of $-\Delta S_M$ at low temperatures.}
\label{6}
\end{figure}

Figures 5(a) and 5(b) show the isothermal magnetizations with field up to 5 T applied in the $ab$ plane and along the $c$ axis, respectively, from 10 to 100 K where data were taken every 2 K around $T_c$. Below $T_c$, the magnetization along the $c$ axis saturates at a relatively low field, however, it increases slowly at low fields in the $ab$ plane and is more harder to saturate, confirming its large magnetic anisotropy and the easy $c$ axis. The magnetic entropy change can be obtained as:\cite{Pecharsky}
\begin{equation}
\Delta S_M(T,H) = \int_0^H [\frac{\partial S(T,H)}{\partial H}]_TdH.
\end{equation}
With the Maxwell's relation $[\frac{\partial S(T,H)}{\partial H}]_T$ = $[\frac{\partial M(T,H)}{\partial T}]_H$, it can be further written as:\cite{Amaral}
\begin{equation}
\Delta S_M(T,H) = \int_0^H [\frac{\partial M(T,H)}{\partial T}]_HdH.
\end{equation}
In the case of magnetization measured at small discrete magnetic field and temperature intervals, $\Delta S_M(T,H)$ could be practically approximated as:
\begin{equation}
\Delta S_M(T,H) = \frac{\int_0^HM(T_{i+1},H)dH-\int_0^HM(T_i,H)dH}{T_{i+1}-T_i}.
\end{equation}
Figures 6(a) and 6(b) gives the calculated $-\Delta S_M$ as a function of temperature in various fields up to 5 T applied in the $ab$ plane and along the $c$ axis, respectively. All the $-\Delta S_M(T,H)$ curves present a pronounced peak around $T_c$, similar to those obtained from heat capacity data [Fig. 4(a)], and the peak broads asymmetrically on both sides with increasing field. The maximum value of $-\Delta S_M$ under 5 T reaches 2.68 J kg$^{-1}$ K$^{-1}$ in the $ab$ plane, which is 37 $\%$ smaller than that of 4.24 J kg$^{-1}$ K$^{-1}$ along the $c$ axis. It is interesting to note that the values of $-\Delta S_M$ for the $ab$ plane are negative at low temperatures in low fields, however, all the values are positive along the $c$ axis, indicating large anisotropy. Fig. 6(c) exhibits the field dependence of $-\Delta S_M$ at low temperatures, in which the sign change is clearly observed with the field in the $ab$ plane but not along the $c$ axis. This originates most likely from the competition of the temperature dependence of magnetic anisotropy and the magnetization. The anisotropy decreases with increasing temperature, whereas the magnetization may exhibit opposite behavior. At low fields, the magnetization at higher temperature could be larger than that at lower temperature [Figs. 1(a,c) and Fig. 5(a)], which gives a negative $-\Delta S_M$. To shed more light on the nature of magnetic transition, we have further analyzed the magnetocaloric data following a recently proposed scaling model.\cite{Franco} It is constructed by normalizing all the $-\Delta S_M$ curves against the respective maximum $-\Delta S_M^{max}$, namely, $\Delta S_M/\Delta S_M^{max}$ by rescaling the temperature $\theta$ below and above $T_c$ as defined in the following equations:
\begin{equation}
\theta_- = (T_{peak}-T)/(T_{r1}-T_{peak}), T<T_{peak},
\end{equation}
\begin{equation}
\theta_+ = (T-T_{peak})/(T_{r2}-T_{peak}), T>T_{peak},
\end{equation}
where $T_{r1}$ and $T_{r2}$ are the temperatures of the two reference points that have been selected as those corresponding to $\Delta S_M(T_{r1},T_{r2}) = \frac{1}{2}\Delta S_M^{max}$. As shown in the insets of Figs. 6(a) and 6(b), all the $-\Delta S_M(T,H)$ curves in various fields collapse into a single curve, indicating a second-order magnetic transition for CrI$_3$.

The rotating magnetic entropy change $\Delta S_M^R$ induced by rotating the applied magnetic field from the $ab$ plane into the $c$ axis direction can be represented as:
\begin{equation}
\begin{aligned}
&\Delta S_M^R(T,H) = S_M(T,H_c)- S_M(T,H_{ab}) \\
&= [S_M(T,H_c)-S_M(T,0)]-[S_M(T,H_{ab})-S_M(T,0)] \\
&= \Delta S_M(T,H_c)-\Delta S_M(T,H_{ab}).
\end{aligned}
\end{equation}
It illustrates that $\Delta S_M^R$ is equal to the difference value of the magnetic entropy change $\Delta S_M$ for $H//c$ and for $H//ab$, as depicted in Fig. 7. The maximum $-\Delta S_M^{Rmax}$ and the width of the main peak increase with increasing field. The field dependence of $-\Delta S_M^{Rmax}$ is presented in the inset of Fig. 7, changing from 0.94 J kg$^{-1}$ K$^{-1}$ for 1 T to 1.56 J kg$^{-1}$ K$^{-1}$ for 5 T. In addition, the temperature corresponding to $-\Delta S_M^{Rmax}$ shows weak field dependence, however, an additional anomaly just below that was also observed in line with the second-stage magnetic ordering at a fixed temperature of 48 K.

\begin{figure}
\centerline{\includegraphics[scale=0.95]{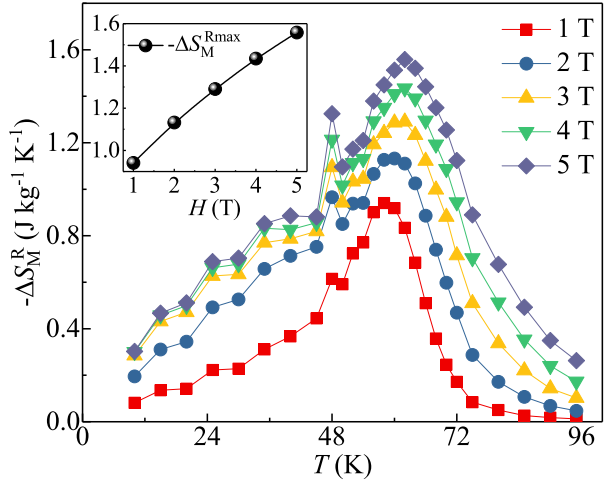}}
\caption{(Color online). Temperature dependence of magnetic entropy change $-\Delta S_M^R$ obtained by rotating from the $ab$ plane to the $c$ axis in various fields. Inset: the maximum $-\Delta S_M^{Rmax}$ as a function of field $H$.}
\label{7}
\end{figure}

\begin{figure}
\centerline{\includegraphics[scale=0.98]{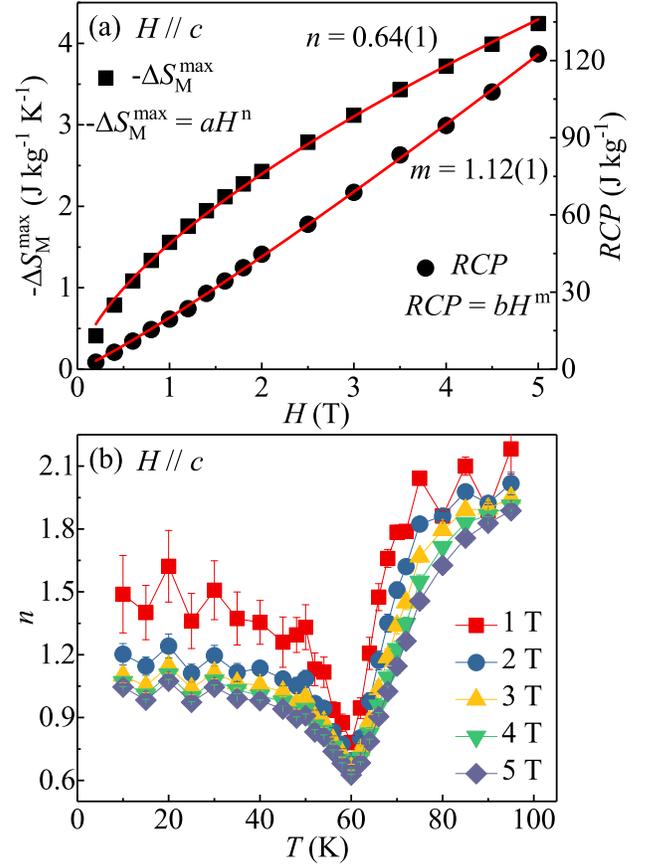}}
\caption{(Color online). (a) Magnetic field dependence of the maximum magnetic entropy change $-\Delta S_M^{max}$ and the relative cooling power RCP with power law fitting in red solid lines. (b) Temperature dependence of $n$ in various fields.}
\label{8}
\end{figure}

\begin{table*}
\caption{\label{tab}Critical exponents of CrI$_3$ compared with various theoretical models.}
\begin{ruledtabular}
\begin{tabular}{lllllllll}
  Material & Theoretical model & Reference & Technique & $\beta$ & $\gamma$ & $\delta$ & $n$ & $m$\\
  \hline
  CrI$_3$ && This work & $-\Delta S_M^{max} = aH^n$  &   &   &   & 0.64(1) & \\
  && This work & $RCP = bH^m$  &   &   &   &   & 1.12(1)\\
  && \cite{LIUYU} & Modified Arrott plot & 0.284(3) & 1.146(11) & 5.04(1) & 0.500(2) & 1.198(1) \\
  && \cite{LIUYU} & Kouvel-Fisher plot & 0.260(4) & 1.136(6) & 5.37(4) & 0.470(1) & 1.186(1) \\
  && \cite{LIUYU} & Critical isotherm  &   &   & 5.32(2) & & 1.188(1) \\
  & Mean field & \cite{Stanley} & Theory & 0.5 & 1.0 & 3.0 & 0.667 & 1.333 \\
  & 3D Heisenberg & \cite{Kaul} & Theory & 0.365 & 1.386 & 4.8 & 0.637 & 1.208 \\
  & 3D XY & \cite{Kaul} & Theory & 0.345 & 1.316 & 4.81 & 0.606 & 1.208 \\
  & 3D Ising & \cite{Kaul} & Theory & 0.325 & 1.24 & 4.82 & 0.569 & 1.207 \\
  & Tricritical mean field & \cite{LeGuillou} & Theory & 0.25 & 1.0 & 5.0 & 0.4 & 1.20

\end{tabular}
\end{ruledtabular}
\end{table*}

For a material displaying a second-order transition,\cite{Oes} the field dependence of the maximum magnetic entropy change shows a power law $-\Delta S_M^{max} = aH^n$,\cite{VFranco} where $a$ is a constant and the exponent $n$ is related to the magnetic order and depends on temperature and field as follows:\cite{Francos}
\begin{equation}
n(T,H) = dln\mid \Delta S_M \mid/dln(H).
\end{equation}
For a FM system above $T_c$, the direct integration of the Curie-Weiss law indicates that $n = 2$. Based on a mean field approach, it becomes field independent at $T_c$ and corresponds to $n = 2/3$.\cite{Oes} However, recent experimental results exhibit deviation from $n = 2/3$ in the case of some soft magnetic amorphous alloys.\cite{VFranco} In addition, there is a relationship between the exponent $n$ at $T_c$ and the critical exponents of the materials as follows:
\begin{equation}
n(T_c) = 1+[\frac{\beta-1}{\beta+\gamma}] = 1+\frac{1}{\delta}[1-\frac{1}{\beta}],
\end{equation}
where $\beta$, $\gamma$, and $\delta$ are related with the spontaneous magnetization $M_s$ below $T_c$, the inverse initial susceptibility $\chi_0^{-1}$ above $T_c$, and the isotherm $M(H)$ at $T_c$, respectively.
In the case of magnetic refrigerators there is another important parameter that evaluates its cooling efficiency of the refrigerant is the relative cooling power (RCP):\cite{Gschneidner}
\begin{equation}
RCP = -\Delta S_M^{max} \times \delta T_{FWHM},
\end{equation}
where $-\Delta S_M^{max}$ is the maximum entropy change near $T_c$ and $\delta T_{FWHM}$ is the full-width at half maxima. The RCP corresponds to the amount of heat that can be transferred between the cold and hot parts of the refrigerator in an ideal thermodynamic cycle.\cite{Tishin} Actually, the RCP also depends on the magnetic field $H$ with the rule $RCP = bH^m$, where $m$ is related to the critical exponent $\delta$ as follows
\begin{equation}
b = 1+\frac{1}{\delta}.
\end{equation}
Figure 8(a) summarized the field dependence of $-\Delta S_M^{max}$ and RCP with $H//c$. The value of RCP is calculated as 122.6 J kg$^{-1}$ for the magnetic field change of 5 T for CrI$_3$, which is about one half of those in manganites and one order of magnitude lower than in ferrites.\cite{Phan, Maalam} Fitting of the $-\Delta S_M^{max}$ gives that $n = 0.64(1)$ [Fig. 8(a)], which deviates the value of $n$ = 0.667 in the mean-field theory and is close to that of $n = 0.637$ in the three-dimensional (3D) Heisenberg model, in line with its localized magnetism nature. Fitting of the RCP generates that $m = 1.12(1)$ [Fig. 8(a)], which is close to the values estimated from the critical exponent $\delta$. The obtained critical exponents of CrI$_3$ as well as the values of different theoretical models are summarized in Table I.\cite{LIUYU, Stanley, Kaul, LeGuillou} Figure 8(b) displays the temperature dependence of $n$ in various fields, giving a precise value of $T_c$ = 60 K. It could be found that with field change of 5 T the value of $n$ is 1.05 and 1.89 far below and above $T_c$, respectively, consistent with the universal law of the $n$ change.\cite{Oes} With decreasing field, the value of $n$ is nearly unchanged at $T_c$ and higher temperatures, however, it shows visible deviation at lower temperatures, which is most likely contributed by its magnetic anisotropy effect.

\section{CONCLUSIONS}

In summary, we have studied in detail the magnetism and magnetocaloric effect of bulk CrI$_3$ single crystal. The second-stage magnetic transition was clearly observed at $T$ = 48 K, just below the Curie temperature $T_c$ = 61 K, indicating two-step magnetic ordering and suggesting that detailed neutron scattering measurements are of interest to shed more light on its microscopic mechanism. A second-order transition from the PM to FM phase around $T_c$ has been confirmed by the scaling analysis of magnetic entropy change $-\Delta S_M$. The $-\Delta S_M$ follows the power law of $H^n$ with $n(T,H) = dln\mid \Delta S_M \mid/dln(H)$, as well as the field dependence of RCP. The $n$ values reach the minimum at 60 K, i.e., its actual $T_c$. Considering its ferromagnetism can be maintained upon exfoliating bulk crystals down to a single layer, further investigation on the size dependence of magnetocaloric effect is of interest.

\section*{Acknowledgements}
This work was supported by the US DOE-BES, Division of Materials Science and Engineering, under Contract No. DE-SC0012704 (BNL).

\end{document}